\documentclass[acmsmall,screen,nonacm]{acmart}

\usepackage{longtable}
\usepackage{array}
\usepackage{xurl}

\newcommand{\rqone}{How do programmers' self-reported activities and experiences differ between programming with and without AI assistance?}
\newcommand{\rqtwo}{How do these differences vary between more and less experienced programmers?}

\AtBeginDocument{%
  }

\begin{document}

\title{Perceived Differences in Activity and Experience Between Developing Software With and Without AI}

\author{Rubash Mali}
\affiliation{%
  \institution{Utah State University}
  \city{Logan}
  \state{Utah}
  \country{USA}}
\email{rubash.mali@usu.edu}

\author{Juho Leinonen}
\affiliation{%
  \institution{Aalto University}
  \city{Espoo}
  \country{Finland}}
\email{juho.2.leinonen@aalto.fi}

\author{John Edwards}
\affiliation{%
  \institution{Utah State University}
  \city{Logan}
  \state{Utah}
  \country{USA}}
\email{john.edwards@usu.edu}

\renewcommand{\shortauthors}{Mali et al.}

\begin{abstract}
The fundamental changes in software engineering driven by generative AI will have profound effects not only on the number of available software engineering jobs but also on what software engineering itself looks like. As the career evolves, questions arise about what the skillset and interests of the next generation of software engineers will be and whether long-time engineers will continue to be attracted to the career. In this paper, we contribute to answering these questions by studying how software development with AI tools is perceived to differ from traditional development and how the differences are modulated by years of programming experience. We find that, in terms of programming activities, engineers report less talking with peers about technical problems, less decomposition, less searching of documentation, and less using a debugger, while they report more task switching. We also find experiential differences, including less deep focus and less sense of control over code and design. We call these the \emph{different craft} results, in contrast to core code-production activities such as writing, refactoring, and fixing code, which show only small mean differences. We also find that these differences are amplified by years of programming experience (an \emph{experience gradient}), with software engineers who started programming before the age of generative AI perceiving greater changes in both activity and developer experience. We discuss our findings and their implications on training of the next generation of engineers and on engineers who may have difficulty transitioning to the AI-infused world of software engineering.
\end{abstract}



\maketitle

\section{Introduction}
Concerns have been raised about generative AI (which we refer to simply as AI in this paper) potentially taking away software engineering jobs. However, AI may also fundamentally change the nature of software engineering and what it means to be a software developer. Changes in the activities involved in programming, e.g. increasingly directing or orchestrating AI rather than performing every part of the work directly, could affect the developer experience. Some changes are positive, such as developers delegating less enjoyable activities, such as writing tests, to AI~\cite{sergeyuk2025using}. But there may be negative effects, such as developers experiencing the flow state less often~\cite{vella2026impact}. AI assistance could thus make programming more appealing in some respects while diminishing other aspects that programmers value. This raises questions about how programmers experience these changes. 

Most recent work approaches the use of AI in software engineering through the lens of what tasks the developer can offload to AI. Developers have been found to use AI in implementing features, writing tests, bug triaging, refactoring, and producing natural-language artifacts~\cite{sergeyuk2025using}. Another study found that engineers use AI beyond automating the production of code, interacting with AI in exploration of different solutions~\cite{barke2023grounded}. Yet the high quality of AI-assistive tools doesn't remove the necessity of understanding, checking, and correcting AI-generated code~\cite{mozannar2024reading}. In a recent longitudinal study, Vella and Blincoe~\cite{vella2026impact} report a perceived shift from creation toward verification, alongside differing trajectories in feedback loops, cognitive load, and flow. 

But there exists a gap in understanding the effect of AI-use on the developer experience broadly in terms of activities, motivation, engagement, satisfaction, and enjoyment, i.e., what is the software developer's perception of engineering with and without AI? And as a corollary, will AI-assisted programming attract people with different interests and motivations compared to engineers of the pre-AI era?

Some prior works provide accounts of task-specific AI use and perceived changes in software development~\cite{sergeyuk2025using,vella2026impact}. We build on these works, expanding the focus beyond activities to the developer experience and obtaining ratings of perceived activity when using and not using AI assistance. In short, we focus on the question: how do programmers characterize programming activities and experiences in programming with and without AI. We also investigate whether the differences vary with programming experience. The latter consideration is motivated by evidence that experience is associated with the roles developers assign to AI. Zakharov et al.~\cite{zakharov2025teacher} found that more experienced developers were more likely to view AI as a junior colleague or content generator whereas less experienced developers were more likely to view it as a teacher. However, differences in these perceived roles do not establish whether the contrast between AI-assisted and non-AI-assisted programming also differs across experience groups at the level of specific activities and experiences. Examining these two contexts together offers a way to investigate this relationship without treating either various programming tasks, or programmers with varying experience levels, as a homogenous category.

In this paper, we investigate these questions through a survey in which programmers provide paired ratings for programming with and without AI assistance. The items in the survey focus on traditional programming activities and the programming experience. The items include questions related to, for example, problem decomposition, writing code from scratch, debugging, control over code design and architecture, code reviews, deep focus, understanding program failures, and manually writing unit tests. For each item, the respondents rated how strongly it applies when programming with AI and when programming without AI. We first examine the differences between respondents' AI and non-AI ratings across the overall sample, and then examine whether these differences vary between more and less experienced programmers. Our analysis focuses on the following two research questions:

\begin{itemize}
    \item[RQ1] \rqone{}
    \item[RQ2] \rqtwo{}
\end{itemize}

This paper makes the following contributions:
\begin{itemize}
    \item We provide evidence that software engineers perceive significant differences in activity and experience between programming with and without AI, including differences in use of documentation, code review, control over code, flow state, and task switching.
    \item We show that these differences are greater in magnitude for programmers with five or more years of experience (those with experience before the availability of generative AI tools).
    \item We discuss the implications of these results on the software engineering community in terms of recruitment and training of new engineers and of satisfaction and retention of engineers transitioning to programming with AI assistance.
\end{itemize}

Together, these contributions provide novel insights into AI-assisted programming beyond code production to the broader activities and experiences involved in software development.

\section{Related Work}
\label{sec:related}

We review prior work on AI-assisted programming, the experience and social side of programming, the individual activities that make up programming, and programming experience and AI use, and then state the gaps our study addresses.

\subsection{AI Assistance and Programming Activities}
\label{sec:related-ai}

Barke et al.~\cite{barke2023grounded} observed 20 programmers using GitHub Copilot and identified two modes: in \emph{acceleration} mode the programmer knows what to do next and uses the assistant to get there faster; in \emph{exploration} mode the programmer is unsure how to proceed and uses it to explore options. Liang et al.~\cite{liang2024survey} surveyed 410 developers about why they use assistants; brainstorming solutions was a weaker motivation than reducing keystrokes, finishing tasks quickly, and recalling syntax, and a main reason for not using assistants was difficulty controlling the tool. These findings need not conflict: Barke et al.\ observed behavior during tasks, whereas Liang et al.\ asked about motivations, so a programmer may explore when unsure how to proceed even if brainstorming is not a primary reason for using the tool.

Vaithilingam et al.~\cite{vaithilingam2022expectation} found that most of the 24 participants in their study preferred Copilot because it gave a useful starting point, which the authors note helps users who are stuck, echoing Barke et al.'s finding that not knowing how to begin leads into exploration~\cite{barke2023grounded}. But participants struggled to understand, edit, and debug generated code, and Mozannar et al.~\cite{mozannar2024reading} quantified such costs: verifying suggestions took 22.4\% of session time, and assistant-specific states together 51.5\%. AI thus adds prompting and verification work rather than simply removing work. Information seeking is also shifting toward AI~\cite{alhaque2024infoseeking}, with Stack Overflow activity declining after ChatGPT's release~\cite{delriochanona2024llm}.

Sergeyuk et al.~\cite{sergeyuk2025using} surveyed 481 programmers about five activities and found that they most wanted to delegate the activities they enjoy least (writing tests and natural-language artifacts), whereas implementing new features was the most enjoyed activity and the one respondents were least willing to delegate; enjoyment and willingness to delegate were negatively correlated for all five activities, significantly so for four. Delegation may thus depend partly on how developers experience an activity, not only on whether AI can perform it, which motivates examining experiences such as deep focus and control over design alongside the activities themselves. Giray et al.~\cite{giray2025adoption} likewise report perceived benefits alongside unreliable output and validation overhead.

Vella and Blincoe~\cite{vella2026impact} surveyed professional software engineers twice, six months apart. Perceived productivity stayed stable (84\% reported improvement at each time point), and matched respondents reported a shift in the balance of effort from creation toward verification. The share of matched respondents rating AI's impact as worse on at least one of three developer-experience dimensions (cognitive load, feedback loops, flow state) nearly doubled, from 14\% to 27\%; the authors describe flow and cognitive load as eroding, although only feedback loops changed significantly in mean ratings (improving). Engineers also described calibrating how far to trust AI output.

\subsection{The Experience and Social Context of Programming}
\label{sec:related-experience-social}

If the activities that make up the programming experience change, so may how engaging, satisfying, or rewarding the work is. Fran\c{c}a et al.~\cite{franca2020motivation} developed the Theory of Motivation and Satisfaction of Software Engineers (TMS-SE) from a multiple case study of four software organizations. It separates work motivation, expressed through engagement (effort) and focus (attention), from job satisfaction, expressed as happiness. AI may change not only what programmers do but also how engaged, focused, and satisfied they are.

In a survey of more than 2,000 developers, the most frequent causes of unhappiness were being stuck in problem solving, time pressure, and bad code quality, with mundane or repetitive tasks and unexplained broken code also in the top ten~\cite{graziotin2017unhappiness}; most high-quality studies of time pressure report higher productivity but lower quality~\cite{kuutila2020time}. Activities targeted by automation are not necessarily activities developers dislike: developers rated finding and fixing bugs as more satisfying than their general work and wanted to stay involved when repair is automated~\cite{winter2023bugfixing}. With the delegation findings above~\cite{sergeyuk2025using}, this raises the possibility that automating tedious work may be experienced differently from automating satisfying work.

Flow~\cite{csikszentmihalyi1990flow} is a state of deep absorption in which the perception of time is transformed, and the deep-focus item in our survey relates to it more directly than to TMS-SE's focus. About 73\% of 684 open-source developers reported losing track of time always or frequently while programming~\cite{lakhani2005hackers}, and developers feel productive when they complete tasks without many context switches~\cite{meyer2014perceptions}. Psychological ownership theory~\cite{pierce2001ownership} holds that people come to feel something is theirs by controlling it, coming to know it intimately, and investing themselves in it. Code-generating LLMs often produce a single point solution by default, obscuring alternative designs and leaving some design decisions implicit~\cite{zamfirescu2025beyond}, which may alter these routes and motivates our items on control over design and on losing track of what code is supposed to do.

Programming is also a social activity. Weinberg~\cite{weinberg1971psychology} advocated \emph{egoless programming}, in which programmers invite colleagues to review their code, and code review at Microsoft yielded knowledge transfer, team awareness, and alternative solutions beyond finding defects~\cite{bacchelli2013expectations}. Professional programmers in pairs outperformed individuals on program-design tasks~\cite{lui2008pairs}, pair-programming dialogue patterns were associated with students' self-efficacy and coding performance~\cite{tan2024dialogue}, and most TA feedback on in-class coding exercises in introductory computer science fell short of the authors' quality standard~\cite{cook2022ta}. If AI assistants answer questions programmers would otherwise take to colleagues, these interactions may change: Houck et al.~\cite{houck2025space} found that only 48\% of AI adopters agreed AI improved their ability to collaborate, and engineering managers reported fewer interruptions as developers relied less on colleagues for quick answers.

\subsection{Individual Programming Activities}
\label{sec:related-activities}

Prior work has examined the activities and experiences in our survey, or closely related constructs, usually without reference to AI: problem decomposition~\cite{ratcliff1985decomposition} and algorithm design~\cite{hundhausen2008algorithms}; pseudo-code and drawings of code~\cite{bellamy1994pseudocode,cherubini2007whiteboard}; debugging with IDE debuggers~\cite{beller2018dichotomy} and asking why a program behaved as it did~\cite{ko2004whyline}; code and pull-request review~\cite{bacchelli2013expectations,bouraffa2025review}; unit testing~\cite{daka2014unit,runeson2006unit}; learning another programming language~\cite{shrestha2020herewego}, applying design patterns~\cite{chatzigeorgiou2008patterns}, and changing paradigms~\cite{miller1995paradigm}; architectural decision-making~\cite{demir2024architectural}; task switching~\cite{coman2008tasks,meyer2014perceptions}; and memory of source code~\cite{kruger2018remember}. Beller et al.~\cite{beller2018dichotomy}, for instance, found that IDE debuggers are used less often than expected, a useful baseline for our debugger result. Table~\ref{tab:rq1-items} marks how each item relates to these sources.
\subsection{Programming Experience and AI}
\label{sec:related-experience}

Zakharov et al.~\cite{zakharov2025teacher} surveyed 3,380 developers and found that experience did not predict AI adoption but was associated with the role developers assigned to AI: more experienced developers more often saw AI as a junior colleague or content generator, or assigned it no role, whereas less experienced developers mainly saw it as a teacher. Al Haque et al.~\cite{alhaque2024infoseeking} stress that foundational knowledge helps developers make use of AI-provided information, especially for learning.

\subsection{Gaps in Prior Work and How Our Study Addresses Them}
\label{sec:related-gaps}

Prior work gives detailed accounts of task-specific AI use and perceived changes in software development~\cite{sergeyuk2025using,vella2026impact}. Two questions remain unresolved by the studies reviewed here, and they correspond to our research questions.

\paragraph{With versus without AI, across a broad set of items.} Controlled studies compare AI-assisted and non-AI programming directly, but on few tasks and outcomes: Peng et al.~\cite{peng2023impact} (two groups, one HTTP-server task, completion time), Vaithilingam et al.~\cite{vaithilingam2022expectation} (Copilot versus IntelliSense, three tasks), and Martin-Lopez et al.~\cite{martinlopez2026more} (69 participants, mostly students, two tasks; time, completeness, and ability to answer questions about their code); Mozannar et al.~\cite{mozannar2024reading} had no non-Copilot baseline. None of these controlled studies rates a broad, common set of activities and experiences under both conditions. Surveys cover more ground but ask about AI use and delegation~\cite{sergeyuk2025using}, perceived effects across the SPACE dimensions~\cite{houck2025space}, or changes in time use and developer experience~\cite{vella2026impact}; the surveys reviewed here do not obtain parallel AI and No-AI ratings for the same individual activities and experiences, so they cannot distinguish an activity rated highly in both contexts from one whose rating differs between them. RQ1 addresses this: each respondent rates the same 26 items with and without AI, and we analyze the paired difference per item.

\paragraph{Dependence on programming experience.} Beyond the role developers assign to AI (Section~\ref{sec:related-experience}), Mozannar et al.~\cite{mozannar2024reading} compared acceptance rates by experience but state that their results are not sufficient to determine its influence. Becker et al.~\cite{becker2025measuring} ran a randomized trial with 16 experienced open-source developers on 246 tasks (allowing AI increased completion time by 19\%) but recruited only experienced developers, and Zare and Assal~\cite{zare2026copilot} compared novice and experienced developers with and without Copilot on code security, where Copilot generally had no effect for either group. The studies reviewed above do not examine whether the within-person AI--No-AI contrast across a broad set of activities and experiences varies with experience. RQ2 does so, comparing programmers with fewer than five and five or more years of experience.

Because the ratings are self-reported and partly retrospective (Section~\ref{sec:threats}), our results map perceived differences, complementing the controlled and observational studies above rather than estimating AI's effects on behavior.

All topics of our survey questions are taken from the literature. Table~\ref{tab:rq1-items} maps each item to its sources and indicates whether each item studies the same activity as the source, studies a related construct, or operationalizes a theory.

\section{Methods}
We conducted a survey among professional software engineers and computer science students. Student respondents were enrolled at a mid-sized R1 university in the United States at the time of the survey. The survey asks questions about each of 26 programming activities and experiences, such as \textit{I use a debugger to step through code line by line} and \textit{I enter a deep focus state where time passes unnoticed}. All 26 questions are listed in Table~\ref{tab:rq1-items}. For each activity, the respondent rates their agreement on a 5-point Likert scale under two conditions: when they are programming with AI and when they are programming without AI. The paired ratings answer RQ1, and background questions let us compare the AI/No-AI differences between more- and less-experienced respondents (RQ2).

\subsection{Survey Instrument}
\textbf{Background items.} Respondents reported their primary role (software engineer, student, researcher, or other with free text), total years of programming experience, years programming without AI, years programming with AI, and weekly programming hours. Response bands for years of programming experience and years of programming with/without AI are 0, 0--1, 1--2, 2--5, 5--10, 10--15, and 15+ years.

\textbf{Activity items.} The 26 activity items are grouped into four sections: core programming activities (Q6--Q14), the programming experience (Q15--Q21), negative programming experiences (Q22--Q26), and traditional programming activities (Q27--Q31).
Table~\ref{tab:rq1-items} lists them by size of the difference between AI and no-AI programming. Each item describes an activity or experience that prior work documents as part of programming, such as problem decomposition~\cite{ratcliff1985decomposition}, debugging~\cite{beller2018dichotomy,ko2004whyline}, unit testing~\cite{daka2014unit,runeson2006unit}, code review~\cite{bacchelli2013expectations,bouraffa2025review}, information seeking~\cite{alhaque2024infoseeking}, and task switching~\cite{coman2008tasks,meyer2014perceptions}, and that prior work on AI-assisted development gives reason to expect AI assistance changes~\cite{barke2023grounded,liang2024survey,sergeyuk2025using,vella2026impact,houck2025space,giray2025adoption}. For example, engineers report spending less time writing code and a shift in focus from creating code to verifying it~\cite{vella2026impact}, and developers report difficulty controlling what AI assistants generate~\cite{liang2024survey}. The experience items also draw on work on software engineers' motivation and satisfaction~\cite{beecham2008motivation,franca2020motivation,hackman1976motivation}, developer affect~\cite{graziotin2014happy,graziotin2017unhappiness}, flow~\cite{csikszentmihalyi1990flow}, motivation in open-source development~\cite{lakhani2005hackers}, psychological ownership~\cite{pierce2001ownership}, and the social nature of programming~\cite{weinberg1971psychology}. The cited work informed the concepts and activities represented by the items rather than their exact wording. Table~\ref{tab:rq1-items} maps every item to the work it draws on and states whether that work studies the same activity or experience, a related construct, or the theory the item operationalizes. Five items are negatively worded: losing track of what a piece of code is supposed to do, being unable to determine why a program is failing, writing code for tasks that feel pointless or disconnected, repeating the same type of programming task many times, and rushing to complete programming work due to deadlines. Respondents rated these items as worded; reverse-scoring was applied during analysis (Section~\ref{sec:dataprep}).

\begin{figure}
    \centering
    \includegraphics[width=0.8\linewidth]{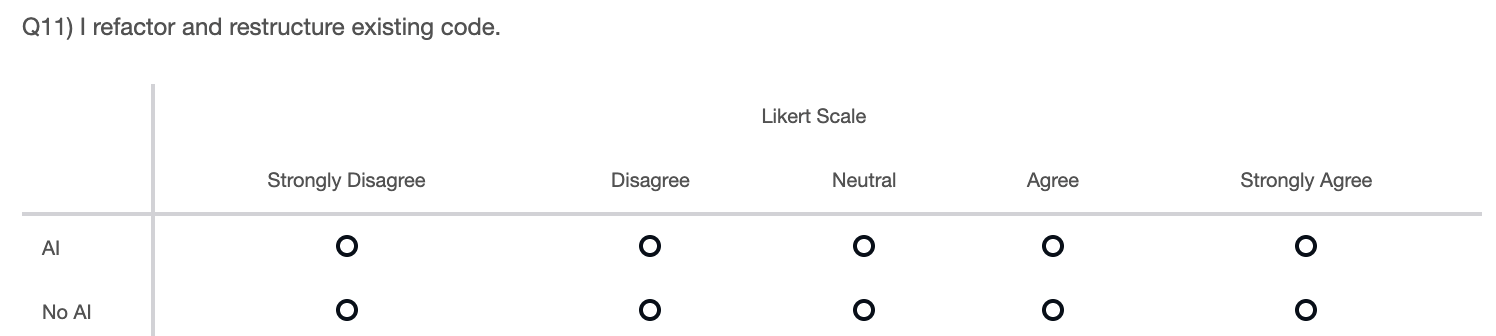}
    \caption{Screenshot of item Q11 of the survey. Survey instructions state that "AI" means "programming using AI" and "No AI" means "programming without AI."}
    \label{fig:q11}
\end{figure}

\textbf{Response format.} For each item, respondents gave two independent ratings on a five-point agreement Likert scale (Strongly Disagree to Strongly Agree): one for programming using AI and one for programming without AI, with the AI row always appearing first. See Figure~\ref{fig:q11} for an example. The instructions state that the two ratings are independent (an activity may be high in both, low in both, or high in only one) and illustrate it with a worked example.

\subsection{Procedure and Participants}
The survey was administered online via Qualtrics between June and September 2026. We recruited respondents through an e-mail to undergraduate and graduate computer science students at a mid-sized U.S.\ university, posts on developer blogging platforms (Hashnode and DEV Community), and programming communities on Reddit, Discord, and Slack. Participation was voluntary and uncompensated, and responses were anonymous; names and e-mail addresses were not collected, and the IP addresses and approximate locations that Qualtrics logs automatically were removed from the dataset before it was shared. The study was determined exempt by our institution's Institutional Review Board and respondents provided informed consent. Because the survey was distributed through open channels, a response rate cannot be computed. 
We received 58 complete responses. In our analysis, we exclude nine respondents who reported no experience programming with AI, because their AI ratings reflect expectation rather than practice (analyzed separately in Section~\ref{sec:nonusers}), and one respondent who completed the survey in 77 seconds (about 1.4 seconds per question across the survey's 57 questions, below our threshold of 2 seconds per question, or about 114 seconds; the next fastest respondent took 126 seconds, while the median was 310 seconds). The analysis sample comprised 48 respondents: 33 software engineers, 9 students, and 6 researchers or educators (Table~\ref{tab:participants}). Respondents who selected ``other'' as their role were reassigned based on their free-text descriptions (technical lead, systems administrator, and returning student with industry experience to software engineer; instructor to researcher/educator).

\begin{table}[]  \centering \caption{Participants (n = 48) by experience group.} \label{tab:participants} \small \begin{tabular}{lrr} \toprule & $<$5 years (n = 14) & $\geq$5 years (n = 34) \\ \midrule \multicolumn{3}{l}{\textit{Role}} \\ \quad Software engineer & 6 & 27 \\ \quad Student & 8 & 1 \\ \quad Researcher/educator & 0 & 6 \\ \addlinespace \multicolumn{3}{l}{\textit{Years programming without AI (Q3)}} \\ \quad 0--2 & 12 & 0 \\ \quad 2--5 & 2 & 9 \\ \quad 5 or more & 0 & 25 \\ \bottomrule \end{tabular} \vspace{2pt} \begin{minipage}{0.8\linewidth} \footnotesize Weekly programming hours: median 25 (IQR 13.8--40; n = 47, one non-numeric response). Years programming with AI: 1--2 years for 40 respondents, 2--5 years for 8. \end{minipage} \end{table} 

\subsection{Data Preparation}
\label{sec:dataprep}

Responses were given on a 1--5 scale. The five negatively worded items were reverse-scored as $6 - \text{score}$.

Reverse-scored items are marked with (R) in all tables and figures. For interpretability, results for these items are discussed using their original wording; for example, a negative AI--No-AI difference for ``losing track of what code is supposed to do'' indicates that respondents associated this experience more strongly with AI-assisted programming.

For each respondent and item, we computed the paired difference as
$\Delta = \text{AI} - \text{No-AI}$. For example, responses of $3$ (AI) and $4$ (no-AI) to the question \textit{I learn new programming languages} would yield a difference score of $3-4=-1$, meaning that the programmer learns new programming languages less when using AI than in the no-AI condition. A negatively framed question, such as \textit{I lose track of what a piece of code is supposed to do} with responses $4$ (AI) and $2$ (no-AI) would have an unadjusted difference score of $4-2=2$, meaning that the programmer loses track of code more often when programming with AI. The adjusted score for analysis comparative with other questions would be $(6-4)-(6-2)=-2$.

There were no missing item ratings and no evidence of straight-lining within either condition. One respondent provided identical ratings across the AI and No-AI conditions for all items. This respondent was retained because the pattern was plausible.

\subsection{Statistical Analysis}
\label{sec:analysis}
\textbf{RQ1.} For each item, we compared AI and No-AI ratings with paired Wilcoxon signed-rank tests, controlling the false discovery rate (FDR) across the 26 items (Benjamini--Hochberg; Holm as a stricter check). Effect sizes are matched-pairs rank-biserial correlations, and mean paired differences are reported with 95\% bootstrap confidence intervals (10{,}000 resamples of respondents; not multiplicity-adjusted). For non-significant items, we used equivalence tests (two one-sided tests, TOST) with a margin of $\pm 0.5$ scale points (half a response category) to distinguish practically unchanged activities from inconclusive ones.

\textbf{RQ2.} We compared less-experienced ($<$5 years, n = 14) and more-experienced ($\geq$5 years, n = 34) programmers; the rationale for the cutoff is given in Section~\ref{sec:rq2}. The primary test was a linear mixed-effects model on all item-level ratings (48 respondents $\times$ 26 items $\times$ 2 conditions): rating $\sim$ condition $\times$ experience, with crossed random intercepts and random condition slopes for respondents and items, so that no items are averaged. We complemented the model with Mann--Whitney comparisons of respondents' mean differences and item-level comparisons with FDR correction. RQ2 was not specified before data collection , so we treat it as exploratory.

\textbf{Robustness and non-users.} Robustness checks (full sample, software engineers only, role adjustment) and the comparison with respondents without AI experience are described together with their results (Sections~\ref{sec:robustness} and~\ref{sec:nonusers}).

\textbf{Software and data.} Analyses were conducted in Python using SciPy, statsmodels (mixed-effects models), and factor\_analyzer. Because an exploratory factor analysis did not support aggregating items into scales (KMO = .60, one weak factor explaining 25\% of variance), we analyze each item individually. Anonymized data, analysis notebooks (including the item-level group comparisons and the factor analysis), the survey instrument, and a table mapping each survey item to the literature it draws on are available at \href{https://osf.io/gkw4v/overview?view_only=a313ac8db7254464a666be4813d1effd} {this link}.

\newlength{\itemw}
\newsavebox{\itemtab}

\section{Results}
\subsection{RQ1: Overall}
\label{sec:rq1}
 
\begin{figure}[]
\centering
\includegraphics[width=\linewidth,height=0.8\textheight,keepaspectratio]{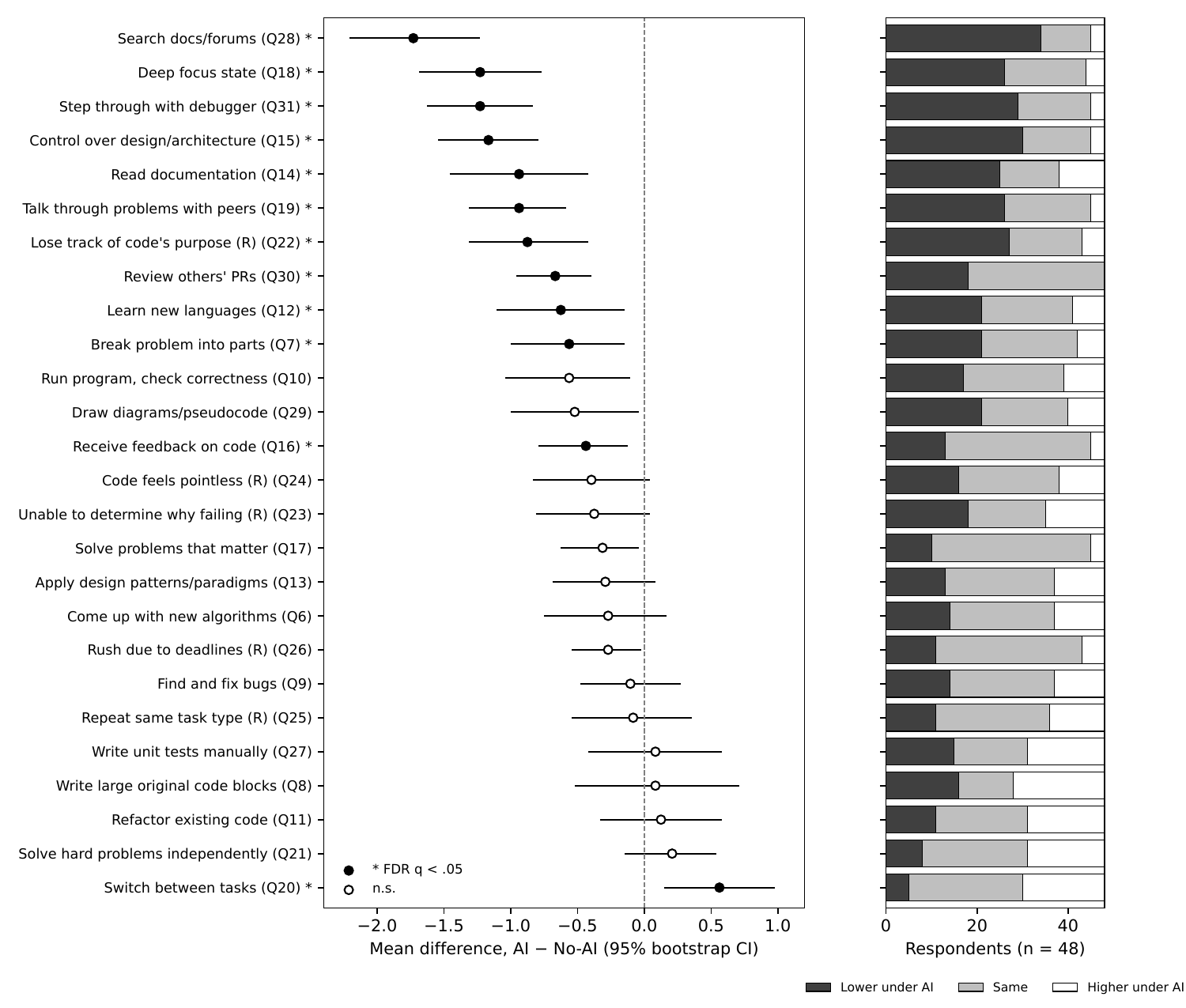}
\caption{AI $-$ No-AI mean differences for all 26 items with 95\% bootstrap confidence intervals (left), sorted by $\Delta$, and the number of respondents rating each item lower, the same, or higher under AI (right; $n = 48$). Asterisks and filled points mark items significant after Benjamini--Hochberg FDR correction ($q < .05$). (R) = reverse-scored; negative values indicate a less favorable rating under AI for every item.}
\label{fig:rq1-forest}
\end{figure}

Of the 26 items, 12 differed significantly between AI and non-AI programming after FDR correction, and 8 of these remained significant under the stricter Holm correction (Table~\ref{tab:rq1-items}, Figure~\ref{fig:rq1-forest}).
Eleven of the twelve were rated lower under AI, most strongly searching documentation or forums ($\Delta = -1.73$), entering a deep focus state ($-1.23$), stepping through code with a debugger ($-1.23$), control over code design and architecture ($-1.17$), reading documentation ($-0.94$), and talking through problems with peers ($-0.94$).
Respondents also reported losing track of what their code is supposed to do more often ($-0.88$; a reverse-scored item, so a negative difference means the experience was more common under AI).
Smaller reductions appeared for reviewing pull requests ($-0.67$), learning new programming languages ($-0.63$), breaking problems into smaller parts ($-0.56$), and receiving feedback on code ($-0.44$).
One item was rated higher under AI: switching between tasks ($+0.56$).
Code-production activities---writing original code, refactoring, finding and fixing bugs, solving hard problems independently, and writing unit tests---showed little or no difference (all $|\Delta| \le 0.21$).
 
\ifdefined\itemw\else\newlength{\itemw}\fi
\ifdefined\itemtab\else\newsavebox{\itemtab}\fi
\providecommand{\lkS}{\ast}
\providecommand{\lkR}{\S}
\providecommand{\lkT}{\#}
\providecommand{\lk}[1]{\ensuremath{^{#1}}}
\begin{table}[]
\centering
\caption{Item-level ratings of programming with and without AI ($n = 48$; 1--5 scale), sorted by $\Delta$. Items are shown in the exact wording of the survey (IDs are the survey question numbers). Symbols after the citations give the item's relation to the cited works (an item can have more than one): $^{*}$\,the item addresses the same activity or experience as a cited work, $^{\S}$\,a cited work studies a related construct, and $^{\ddagger}$\,the item operationalizes a theory defined in a cited work. $\Delta$ = mean paired difference (AI $-$ No-AI) with 95\% bootstrap CI; $r_{\mathrm{rb}}$ = matched-pairs rank-biserial correlation; L/S/H = number of respondents rating the item lower / the same / higher under AI; $q$ = Benjamini--Hochberg FDR-adjusted $p$ from paired Wilcoxon signed-rank tests (bold: $q < .05$, 12/26). $^{\dagger}$Also significant after Holm correction (8/26). (R) = reverse-scored, so higher = less negative experience.}
\label{tab:rq1-items}
\footnotesize
\setlength{\tabcolsep}{3pt}
\renewcommand{\arraystretch}{1.1}
\setlength{\itemw}{2.5in}
\sbox{\itemtab}{%
\begin{tabular}{llcccccr}
\toprule
ID & Survey item & AI $M$ & No-AI $M$ & $\Delta$\;[95\% CI] & $r_{\mathrm{rb}}$ & L/S/H & $q$ \\
\midrule
Q28 & \parbox[t]{\itemw}{\raggedright I search documentation or forums (e.g., Stack Overflow)~\cite{alhaque2024infoseeking,mozannar2024reading}$^{*}$\strut} & 2.46 & 4.19 & $-1.73\;[-2.21,\,-1.23]$ & $-0.86$ & 34/11/3 & \textbf{$<$.001}$^{\dagger}$ \\
Q18 & \parbox[t]{\itemw}{\raggedright I enter a deep focus state where time passes unnoticed~\cite{csikszentmihalyi1990flow,lakhani2005hackers}$^{*\ddagger}$\strut} & 3.02 & 4.25 & $-1.23\;[-1.69,\,-0.77]$ & $-0.87$ & 26/18/4 & \textbf{$<$.001}$^{\dagger}$ \\
Q31 & \parbox[t]{\itemw}{\raggedright I use a debugger to step through code line by line~\cite{beller2018dichotomy}$^{*}$\strut} & 2.69 & 3.92 & $-1.23\;[-1.63,\,-0.83]$ & $-0.90$ & 29/16/3 & \textbf{$<$.001}$^{\dagger}$ \\
Q15 & \parbox[t]{\itemw}{\raggedright I have control over the code design and architecture~\cite{demir2024architectural,pierce2001ownership}$^{\S\ddagger}$\strut} & 3.38 & 4.54 & $-1.17\;[-1.54,\,-0.79]$ & $-0.88$ & 30/15/3 & \textbf{$<$.001}$^{\dagger}$ \\
Q14 & \parbox[t]{\itemw}{\raggedright I read documentation to solve a problem~\cite{alhaque2024infoseeking,mozannar2024reading}$^{*}$\strut} & 3.31 & 4.25 & $-0.94\;[-1.46,\,-0.42]$ & $-0.62$ & 25/13/10 & \textbf{.004}$^{\dagger}$ \\
Q19 & \parbox[t]{\itemw}{\raggedright I talk through technical problems with peers or colleagues~\cite{tan2024dialogue}$^{*}$\strut} & 3.06 & 4.00 & $-0.94\;[-1.31,\,-0.58]$ & $-0.85$ & 26/19/3 & \textbf{$<$.001}$^{\dagger}$ \\
Q22 (R) & \parbox[t]{\itemw}{\raggedright I lose track of what a piece of code is supposed to do~\cite{kruger2018remember,pierce2001ownership}$^{\S\ddagger}$\strut} & 2.27 & 3.15 & $-0.88\;[-1.31,\,-0.42]$ & $-0.68$ & 27/16/5 & \textbf{.003}$^{\dagger}$ \\
Q30 & \parbox[t]{\itemw}{\raggedright I review other developers' code in pull requests~\cite{bacchelli2013expectations,bouraffa2025review}$^{*}$\strut} & 3.25 & 3.92 & $-0.67\;[-0.96,\,-0.40]$ & $-1.00$ & 18/30/0 & \textbf{$<$.001}$^{\dagger}$ \\
Q12 & \parbox[t]{\itemw}{\raggedright I learn new programming languages~\cite{shrestha2020herewego}$^{*}$\strut} & 3.10 & 3.73 & $-0.63\;[-1.10,\,-0.15]$ & $-0.49$ & 21/20/7 & \textbf{.048} \\
Q7 & \parbox[t]{\itemw}{\raggedright I break a large programming problem into smaller parts before writing code~\cite{ratcliff1985decomposition}$^{*}$\strut} & 3.67 & 4.23 & $-0.56\;[-1.00,\,-0.15]$ & $-0.51$ & 21/21/6 & \textbf{.048} \\
Q10 & \parbox[t]{\itemw}{\raggedright I run a program and check its correctness~\cite{mozannar2024reading}$^{*}$\strut} & 3.85 & 4.42 & $-0.56\;[-1.04,\,-0.10]$ & $-0.47$ & 17/22/9 & .064 \\
Q29 & \parbox[t]{\itemw}{\raggedright I draw diagrams or write pseudocode~\cite{bellamy1994pseudocode,cherubini2007whiteboard}$^{*}$\strut} & 3.02 & 3.54 & $-0.52\;[-1.00,\,-0.04]$ & $-0.42$ & 21/19/8 & .077 \\
Q16 & \parbox[t]{\itemw}{\raggedright I receive feedback from instructors/TAs/supervisors regarding code~\cite{bacchelli2013expectations,cook2022ta}$^{*\S}$\strut} & 2.77 & 3.21 & $-0.44\;[-0.79,\,-0.13]$ & $-0.68$ & 13/32/3 & \textbf{.043} \\
Q24 (R) & \parbox[t]{\itemw}{\raggedright I write code for tasks that feel pointless or disconnected~\cite{beecham2008motivation,hackman1976motivation}$^{\S\ddagger}$\strut} & 3.06 & 3.46 & $-0.40\;[-0.83,\,+0.04]$ & $-0.39$ & 16/22/10 & .113 \\
Q23 (R) & \parbox[t]{\itemw}{\raggedright I am unable to determine why a program is failing~\cite{graziotin2017unhappiness,ko2004whyline}$^{*}$\strut} & 3.00 & 3.38 & $-0.38\;[-0.81,\,+0.04]$ & $-0.34$ & 18/17/13 & .127 \\
Q17 & \parbox[t]{\itemw}{\raggedright I solve problems that directly matter to users or to a project~\cite{beecham2008motivation,hackman1976motivation,lakhani2005hackers}$^{*\ddagger}$\strut} & 4.00 & 4.31 & $-0.31\;[-0.63,\,-0.04]$ & $-0.59$ & 10/35/3 & .085 \\
Q13 & \parbox[t]{\itemw}{\raggedright I apply new design patterns and/or programming paradigms~\cite{chatzigeorgiou2008patterns,miller1995paradigm}$^{*}$\strut} & 3.40 & 3.69 & $-0.29\;[-0.69,\,+0.08]$ & $-0.40$ & 13/24/11 & .113 \\
Q6 & \parbox[t]{\itemw}{\raggedright I come up with new algorithms~\cite{hundhausen2008algorithms,lui2008pairs}$^{*}$\strut} & 3.21 & 3.48 & $-0.27\;[-0.75,\,+0.17]$ & $-0.20$ & 14/23/11 & .453 \\
Q26 (R) & \parbox[t]{\itemw}{\raggedright I rush to complete programming work due to deadlines~\cite{graziotin2017unhappiness,kuutila2020time}$^{*}$\strut} & 2.75 & 3.02 & $-0.27\;[-0.54,\,-0.02]$ & $-0.56$ & 11/32/5 & .077 \\
Q9 & \parbox[t]{\itemw}{\raggedright I find and fix bugs in code~\cite{winter2023bugfixing}$^{*}$\strut} & 4.04 & 4.15 & $-0.10\;[-0.48,\,+0.27]$ & $-0.13$ & 14/23/11 & .637 \\
Q25 (R) & \parbox[t]{\itemw}{\raggedright I repeat the same type of programming task many times~\cite{graziotin2017unhappiness}$^{*}$\strut} & 2.77 & 2.85 & $-0.08\;[-0.54,\,+0.35]$ & $-0.09$ & 11/25/12 & .740 \\
Q27 & \parbox[t]{\itemw}{\raggedright I create unit tests manually to verify correctness~\cite{daka2014unit,runeson2006unit}$^{*}$\strut} & 3.48 & 3.40 & $+0.08\;[-0.42,\,+0.58]$ & $+0.08$ & 15/16/17 & .740 \\
Q8 & \parbox[t]{\itemw}{\raggedright I write large blocks of original code from scratch~\cite{mozannar2024reading,sergeyuk2025using}$^{\S}$\strut} & 3.52 & 3.44 & $+0.08\;[-0.52,\,+0.71]$ & $+0.04$ & 16/12/20 & .818 \\
Q11 & \parbox[t]{\itemw}{\raggedright I refactor and restructure existing code~\cite{sergeyuk2025using}$^{*}$\strut} & 4.04 & 3.92 & $+0.13\;[-0.33,\,+0.58]$ & $+0.18$ & 11/20/17 & .470 \\
Q21 & \parbox[t]{\itemw}{\raggedright I solve challenging programming problems independently without any interaction with other humans~\cite{lui2008pairs,weinberg1971psychology}$^{*}$\strut} & 4.06 & 3.85 & $+0.21\;[-0.15,\,+0.54]$ & $+0.34$ & 8/23/17 & .158 \\
Q20 & \parbox[t]{\itemw}{\raggedright I switch between different programming tasks in one session~\cite{coman2008tasks,meyer2014perceptions}$^{*}$\strut} & 3.96 & 3.40 & $+0.56\;[+0.15,\,+0.98]$ & $+0.54$ & 5/25/18 & \textbf{.048} \\
\bottomrule
\end{tabular}}
\ifdim\wd\itemtab>\linewidth\resizebox{\linewidth}{!}{\usebox{\itemtab}}\else\usebox{\itemtab}\fi
\end{table}

\paragraph{Practically unchanged activities.}
Equivalence tests on the 14 non-significant items (Table~\ref{tab:tost}) indicated that four activities were practically unchanged under AI: finding and fixing bugs, repeating the same type of task, rushing due to deadlines, and solving hard problems independently. The last two sat right at the threshold ($p_{\mathrm{TOST}} = .049$ and $.0497$).
Two further items were borderline (writing unit tests, refactoring).
The remaining eight items were inconclusive ($p_{\mathrm{TOST}} \ge .100$).

\begin{table}[]
\centering
\caption{Equivalence tests (TOST, margin $\pm 0.5$ scale points) for the six non-significant items with the lowest $p_{\mathrm{TOST}}$ ($n = 48$). An item is equivalent when $p_{\mathrm{TOST}} < .05$; borderline = $p_{\mathrm{TOST}}$ just above .05. The other eight non-significant items had $p_{\mathrm{TOST}} \ge .100$ (inconclusive: neither significant nor shown to be equivalent). (R) = reverse-scored.}
\label{tab:tost}
\footnotesize
\setlength{\tabcolsep}{4pt}
\begin{tabular}{lccl}
\toprule
Item & $\Delta$ & $p_{\mathrm{TOST}}$ & Conclusion \\
\midrule
Find and fix bugs (Q9) & $-0.10$ & .021 & Equivalent \\
Repeat same task type (R) (Q25) & $-0.08$ & .038 & Equivalent \\
Rush due to deadlines (R) (Q26) & $-0.27$ & .049 & Equivalent \\
Solve hard problems alone (Q21) & $+0.21$ & .0497 & Equivalent \\
Write unit tests manually (Q27) & $+0.08$ & .054 & Borderline \\
Refactor existing code (Q11) & $+0.13$ & .061 & Borderline \\
\bottomrule
\end{tabular}
\end{table}

\subsection{RQ2: Does the AI vs.\ No-AI Difference Depend on Programming Experience?}
\label{sec:rq2}
 
\subsubsection{Experience groups and the five-year cutoff}
 
Respondents were grouped by total programming experience (Q2) into less-experienced ($<$5 years, n = 14) and more-experienced ($\geq$5 years, n = 34) programmers. See Table~\ref{tab:rq2-q3}. 12 of 14 less-experienced respondents reported two years or less of programming without AI, whereas all of the more-experienced respondents had at least two years of experience programming without AI and 25 of 34 reported five years or more. At the time of data collection (June--September 2026), five years approximately corresponds to the arrival of AI coding assistants (GitHub Copilot's preview in mid-2021, followed by ChatGPT in late 2022), and it is one of the boundaries between Q2's predefined response bands.
 
\begin{table}[]
\centering
\caption{Years of programming without AI by experience group. "yrs exp" indicates "total years of programming experience."}
\label{tab:rq2-q3}
\footnotesize
\setlength{\tabcolsep}{3pt}
\begin{tabular}{lrr}
\toprule
Years without AI (Q3) & $<$5 yrs exp & $\geq$5 yrs exp \\
\midrule
0--2 (0 or 1--2) & \textbf{12} & 0 \\
2--5 & 2 & 9 \\
5 or more (5--10, 10--15, 15+) & 0 & \textbf{25} \\
\bottomrule
\end{tabular}
\end{table}
 
\textbf{Sensitivity to the cutoff.} Treating experience as the survey's ordered bands (five are represented in the sample, 1--2 to 15+ years) instead of a cutoff yielded an effect in the same direction: greater experience was associated with a larger AI $-$ No-AI difference (Spearman $\rho = -.34$, $p = .016$; mixed-model trend $-0.25$ per band, 95\% CI [$-0.41$, $-0.09$], $p = .002$; bands treated as equally spaced). Band means were nearly monotonic (1--2: $+0.41$, n = 3; 2--5: $-0.24$, n = 11; 5--10: $-0.56$, n = 16; 10--15: $-0.55$, n = 10; 15+: $-0.92$, n = 8).
 
\subsubsection{Main result}
 
\begin{table}[]
\centering
\caption{AI and No-AI ratings by experience group (respondent-level means across all 26 items; Q22--Q26 reverse-scored). Within-group $p$: paired Wilcoxon; between-group $p$: Mann--Whitney; CI from 10{,}000 bootstrap resamples within groups. Gap = $\Delta_{<5} - \Delta_{\geq5}$ (positive = more negative AI $-$ No-AI difference for $\geq$5 years).}
\label{tab:rq2-main}
\small
\begin{tabular}{lrrrcr}
\toprule
 & $<$5 yrs & $\geq$5 yrs & Difference & 95\% CI & $p$ \\
\midrule
No-AI mean & 3.56 & 3.84 & +0.28 & & .037 \\
AI mean & 3.46 & 3.20 & $-$0.26 & & .312 \\
$\Delta$ (AI $-$ No-AI) & $-$0.10 & $-$0.64 & gap 0.54 & [0.14, 0.94] & .022 \\
Within-group $p$ ($\Delta \neq 0$) & .610 & $<$.001 & & & \\
\bottomrule
\end{tabular}
\end{table}
 
\begin{table}[]
\centering
\caption{Linear mixed-effects model on all 2{,}496 ratings (48 respondents $\times$ 26 items $\times$ 2 conditions): rating $\sim$ condition $\times$ experience, with crossed random intercepts and random condition slopes for respondents and items (uncorrelated variance components, REML). Reference: $<$5 years, No-AI.}
\label{tab:rq2-model}
\small
\begin{tabular}{llrcr}
\toprule
Term & Interpretation & Estimate & 95\% CI & $p$ \\
\midrule
Intercept & $<$5 yrs, No-AI & 3.56 & [3.31, 3.80] & $<$.001 \\
Condition (AI) & AI effect, $<$5 yrs & $-$0.10 & [$-$0.49, 0.29] & .617 \\
Experience ($\geq$5 yrs) & Group difference under No-AI & +0.28 & [0.06, 0.50] & .013 \\
Condition $\times$ Experience & Additional AI effect, $\geq$5 yrs & \textbf{$-$0.54} & [$-$0.95, $-$0.13] & \textbf{.010} \\
\bottomrule
\end{tabular}
 
\vspace{2pt}
\begin{minipage}{0.9\linewidth}
\footnotesize
Variance components: respondent intercept 0.078, respondent AI slope 0.345, item intercept 0.169, item AI slope 0.200; residual 1.23.
\end{minipage}
\end{table}
 
More-experienced programmers showed a larger AI $-$ No-AI difference than less-experienced programmers (condition $\times$ experience $= -0.54$, 95\% CI [$-0.95$, $-0.13$], $p = .010$; Table~\ref{tab:rq2-model}), consistent with the respondent-level comparison ($\Delta = -0.64$ vs.\ $-0.10$; gap 0.54, 95\% CI [0.14, 0.94], Mann--Whitney $p = .022$, $r = .43$; Table~\ref{tab:rq2-main}). Within groups, the difference was clear for more-experienced ($p < .001$) but not detected for less-experienced programmers ($p = .610$). Descriptively, the gap reflected both sides of the comparison: more-experienced programmers rated non-AI programming higher (3.84 vs.\ 3.56; model estimate +0.28, $p = .013$) and, to a similar degree, AI programming lower (3.20 vs.\ 3.46), although the latter difference was not detected on its own ($p = .312$; this does not establish equivalence).

\subsubsection{Item-level pattern}
\begin{figure}[]
\centering
\includegraphics[width=\linewidth,height=0.66\textheight,keepaspectratio]{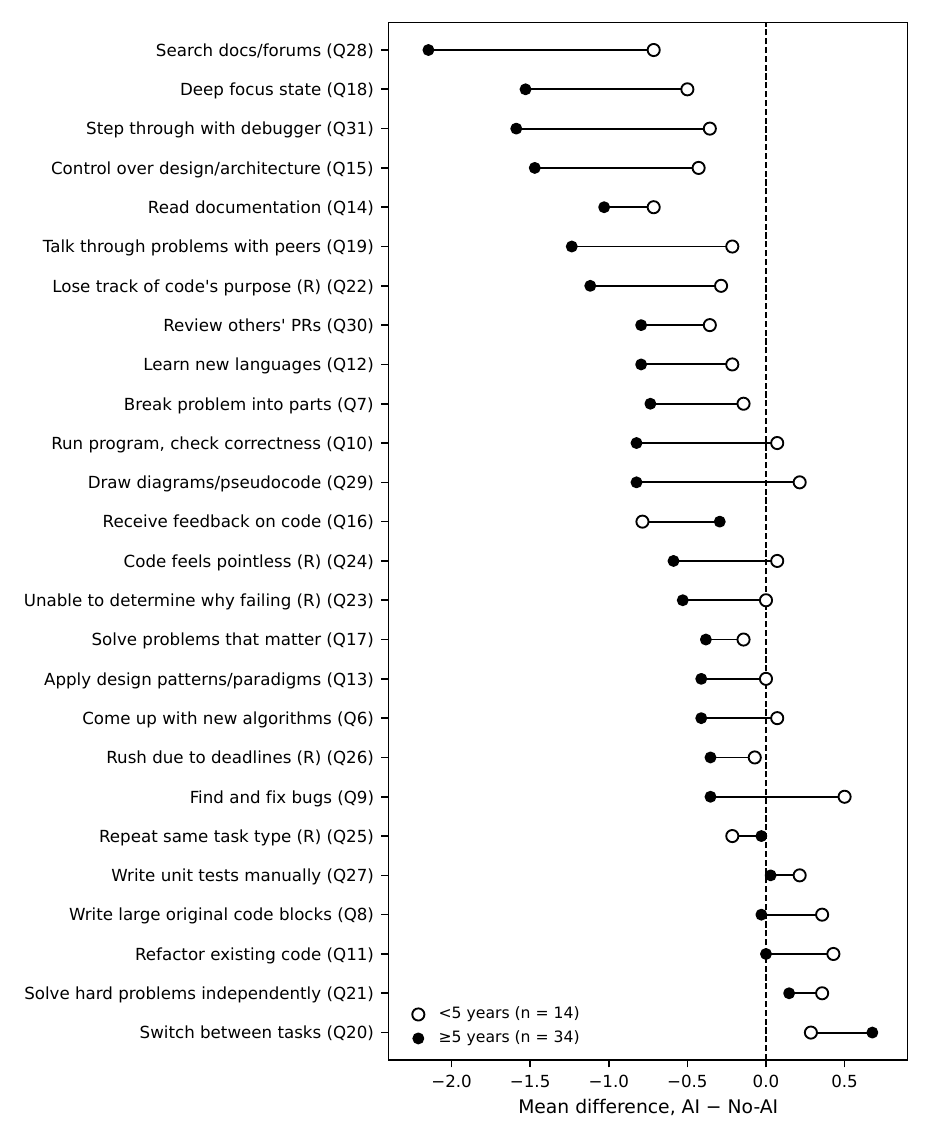}
\caption{Mean AI $-$ No-AI difference by experience group for all 26 items, in the order of Figure~\ref{fig:rq1-forest} (largest overall drop first). Open circles: $<$5 years (n = 14); filled circles: $\geq$5 years (n = 34); each line connects the two groups' differences for one item. The difference is more negative for the $\geq$5-year group on 23 of 26 items, and item-level differences are positively correlated between groups ($\rho = .66$). No individual item difference between groups survives FDR correction. (R) = reverse-scored; negative values indicate a less favorable rating under AI.}
\label{fig:rq2-items}
\end{figure}

More-experienced programmers rated 22 of 26 items lower under AI, compared with 14 of 26 for less-experienced programmers, and 17 of 26 items shifted in the same direction in both groups (Figure~\ref{fig:rq2-items}). The AI $-$ No-AI difference was more negative for the $\geq$5-year group on 23 of 26 items, and item-level differences were positively correlated between groups (Spearman $\rho = .66$, $p < .001$): the two groups showed broadly the same pattern, larger among more-experienced programmers.

\begin{figure}[]
\centering
\includegraphics[width=\linewidth,height=0.75\textheight,keepaspectratio]{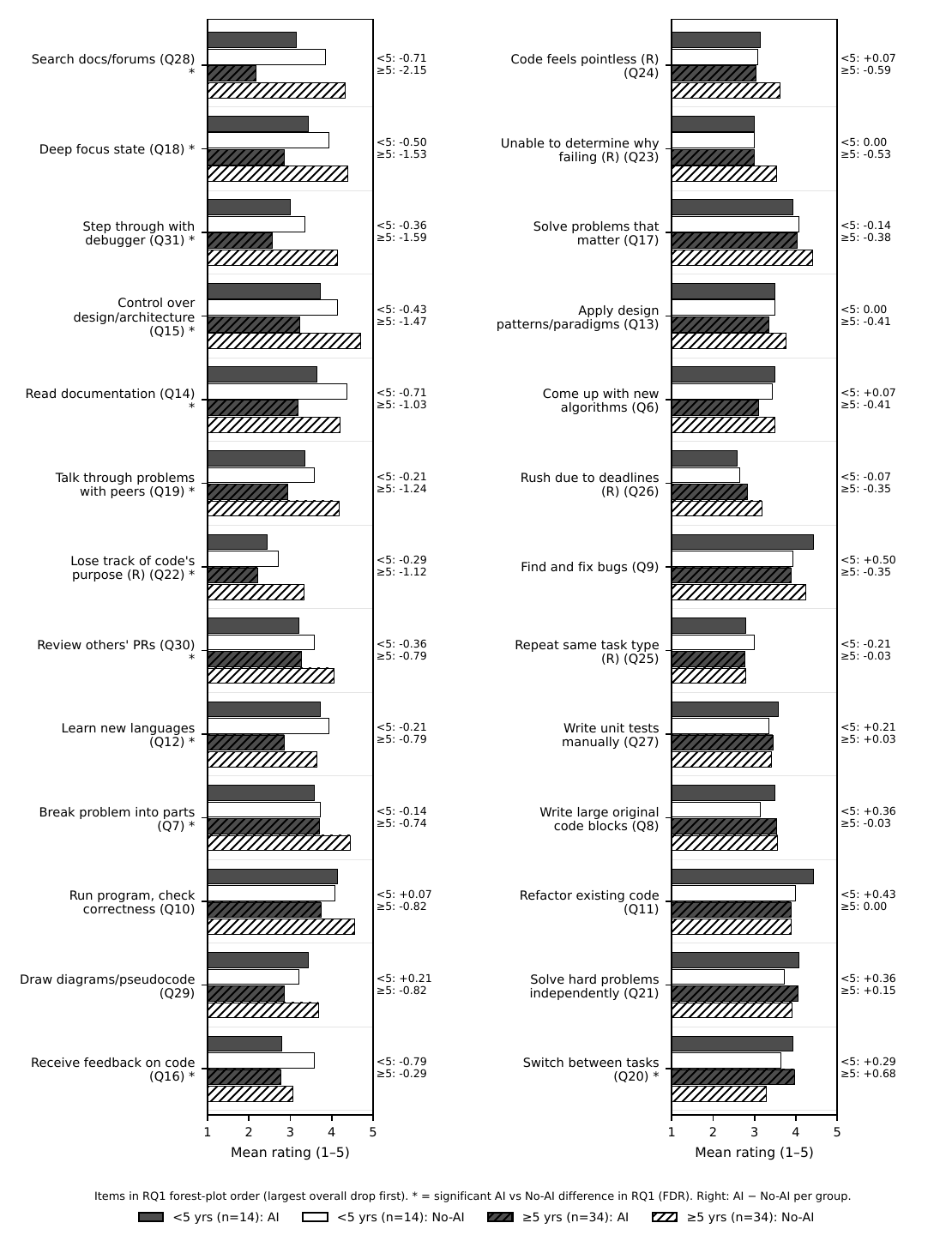}
\caption{AI and No-AI mean ratings (1--5) for all 26 items by experience group (solid: $<$5 years, n = 14; hatched: $\geq$5 years, n = 34; dark: AI; white: No-AI). Values at right: AI $-$ No-AI difference per group. * = significant AI vs.\ No-AI difference in RQ1 (FDR). (R) = reverse-scored; higher = more favorable for all items.}
\label{fig:rq2-bars}
\end{figure}

The three exceptions to the more negative difference among more-experienced programmers were receiving feedback on code ($-0.79$ vs.\ $-0.29$), repeating the same type of task ($-0.21$ vs.\ $-0.03$), and switching between tasks ($+0.29$ vs.\ $+0.68$). On the nine items whose direction differed between groups, the less-experienced group's difference was zero or positive (up to $+0.50$, finding and fixing bugs) while the more-experienced group's was negative or zero. Figure~\ref{fig:rq2-bars} shows the underlying AI and No-AI means.

\newsavebox{\robtab}

\subsection{Robustness Checks}
\label{sec:robustness}
We repeated the main analyses on alternative samples. Without exclusions, 11 of the 12 RQ1 items stayed significant (all but task switching); the four added items came from the non-users (Section~\ref{sec:nonusers}). Among software engineers alone, 8 of the 12 stayed significant and 25 of 26 items kept their direction ($\rho = .97$). The RQ2 interaction held in every check. Excluding the respondent with identical ratings did not change RQ1.

\subsection{Respondents Without AI Experience}
\label{sec:nonusers}

Nine respondents reported no experience programming with AI (three software engineers, three students, one researcher/educator, and two who selected ``other''; five with $<$5 and four with $\geq$5 years of programming experience). All of them had only ever programmed without AI. Because their AI ratings reflect expectation rather than practice, we excluded them from the main analyses and compare them here with the 48 AI users (Table~\ref{tab:nonusers}). With $n = 9$, these results are descriptive.

Non-users gave almost identical mean ratings for programming without AI compared to users (M = 3.76 in both groups; Mann--Whitney $p = .599$) but rated AI-assisted programming much lower (2.50 vs.\ 3.27, $p < .001$). Their AI $-$ No-AI difference was therefore larger ($\Delta = -1.27$ vs.\ $-0.48$; Mann--Whitney $p = .020$, $r = -.49$), and 8 of 9 rated AI lower overall (Wilcoxon $p = .012$). In a mixed-effects model on all 57 respondents' ratings, non-users showed an additional AI drop of $-0.79$ (95\% CI [$-1.30$, $-0.28$], $p = .003$) beyond the users' own AI effect ($-0.48$, $p < .001$), while the groups did not differ under No-AI ($+0.01$, $p = .954$): a group difference was detected under AI but not under No-AI.

At the item level, non-users rated 24 of 26 items lower under AI and had the larger drop on 22 of 26 items. Their item pattern resembled the users' (Spearman $\rho = .54$, $p = .004$; same direction on 19 of 26 items), but the drops were much larger. The clearest divergence was in code production: non-users expected declines in writing large blocks of original code ($-1.67$ vs.\ $+0.08$ for users), finding and fixing bugs ($-1.78$ vs.\ $-0.10$), refactoring ($-1.22$ vs.\ $+0.13$), and solving hard problems independently ($-1.22$ vs.\ $+0.21$), activities that users report as essentially unchanged, and they expected task switching to fall ($-0.67$) whereas users report it rising ($+0.56$). Nine items differed within non-users at unadjusted $p < .05$, but none survived FDR correction. Among non-users, the five with $<$5 years of experience showed a mean $\Delta$ of $-0.78$ and the four with $\geq$5 years $-1.88$, the same direction as in RQ2 but too few respondents to test. Non-users may also differ from users in ways we did not measure, for example by avoiding AI because they dislike it or by using it outside programming.

\begin{table}[]
\centering
\caption{Non-users ($n = 9$) compared with AI users ($n = 48$). (a) Respondent-level means across the 26 items (Q22--Q26 reverse-scored); $\Delta$ = AI $-$ No-AI; difference = non-users $-$ users; $p$: Mann--Whitney. (b) Items on which non-users' expectations diverge most from users' reported experience; L/S/H = non-users rating the item lower / the same / higher under AI. No item-level difference within non-users survives FDR correction.}
\label{tab:nonusers}
\footnotesize
\setlength{\tabcolsep}{4pt}
\begin{tabular}{lcccr}
\toprule
\multicolumn{5}{l}{\textit{(a) Overall}} \\
 & Non-users & Users & Difference & $p$ \\
\midrule
No-AI mean & 3.76 & 3.76 & +0.01 & .599 \\
AI mean & 2.50 & 3.27 & $-$0.78 & $<$.001 \\
$\Delta$ (AI $-$ No-AI) & $-$1.27 & $-$0.48 & $-$0.79 & .020 \\
Rated AI lower overall & 8 of 9 & 30 of 48 & & \\
\midrule
\multicolumn{5}{l}{\textit{(b) Selected items}} \\
Item & $\Delta$ non-users & $\Delta$ users & L/S/H non-users & \\
\midrule
Write large original code blocks (Q8) & $-$1.67 & +0.08 & 7/1/1 & \\
Find and fix bugs (Q9) & $-$1.78 & $-$0.10 & 6/2/1 & \\
Refactor existing code (Q11) & $-$1.22 & +0.13 & 6/2/1 & \\
Solve hard problems alone (Q21) & $-$1.22 & +0.21 & 4/4/1 & \\
Switch between tasks (Q20) & $-$0.67 & +0.56 & 4/3/2 & \\
\bottomrule
\end{tabular}
\end{table} 
 
\section{Discussion}

\subsection{RQ1: \rqone{}}
It is clear from the results that programming with AI is different than programming without AI, at least from the engineer's perspective. Across a variety of activities and experiences, AI has effected significant changes in what a software engineer does and experiences while designing, writing, testing, and debugging code. We refer to these as the \emph{different craft} results: many of the activities and experiences surrounding code production are rated lower with AI, while core code-production activities such as writing, refactoring, and fixing code show only small mean differences. In this section, we discuss programming activities and experiences that are significantly more/less present when using/not using AI, respectively. See Table~\ref{tab:rq1-items}.

\subsubsection{I break a large programming problem into smaller parts before writing code; I use a debugger to step through code line by line.} \label{sec:less-debugger} Engineers decompose less and use the debugger less when using AI. These are both tasks specifically targeted by AI tools. Skills in both decomposition and debugging are valued in the software engineering industry with computer science educational programs targeting both~\cite{margulieux2016employing,yang2025decoding} and decomposition taking a prominent spot in computational thinking~\cite{wing2006computational}.

The result from these two questions underscores the importance that training programs, including computer science bachelor programs, be responsive to changing needs of the industry. It also serves as a call to the research community to identify exactly what software engineers need to know. For example, while our results suggest that decomposition is done less, we don't know whether decomposition is an important skill when directing AI in its work.

\subsubsection{I search documentation or forums (e.g., Stack Overflow); I read documentation to solve a problem.} Engineers read documentation less often when using AI. This likely has a combination of two explanations. The first is that engineers can now ask AI for information instead of reading documentation. The second is that engineers can rely less on learning APIs and language features since AI is now doing the work. This is an expected consequence of programming with AI; by definition, AI is doing the work, which includes making library calls and using a language appropriately. In many ways, writing software traditionally is an educational experience, learning new APIs and language features. But in the age of AI, as confirmed by our survey, it is unclear what learning is being done by engineers, beyond learning how to more effectively interact with the AI.

\subsubsection{I learn new programming languages.} Related to the easing of reliance on documentation is the fact that engineers are either learning fewer languages or learning new languages more superficially. A fundamental skill of marketable software engineers in the past has been learning new technologies quickly, including languages. Instead of learning new technologies, engineers are improving their craft in the ability to architect software and direct AI agents efficiently. These are very different skills than learning a programming language, moving from mathematics-related skills of using language schema to dynamic management skills more related to ideas found in the business world.

\subsubsection{I review other developers’ code in pull requests; I receive feedback from instructors/TAs/supervisors
regarding code.}
\label{sec:results-trust}
Both of these questions are related to putting more trust in AI-generated code than in the code produced by humans. Much discussion has been made about software engineers needing more skill in reading code than in writing code, but this result suggests that reading code is also becoming less relevant.

\subsubsection{I talk through technical problems with peers or colleagues.} The question does not specify whether the technical questions are low-level or high-level, but it appears that discussions of both are becoming less common. What does this mean for social interaction in the workplace? It doesn't say anything about purely social interaction, but it does say something about conversations that start as technical discussions. Our survey did not directly measure experiential aspects of workplace interaction, but we can infer that they are being affected by use of AI.

\subsubsection{I enter a deep focus state where time passes unnoticed; I switch between different programming tasks in one session.} The fact that engineers report entering deep focus less when using AI is critical in our efforts to understand what software engineering with AI looks like. Deep focus is associated with swift passage of time and happiness~\cite{csikszentmihalyi1990flow,Csikszentmihalyi2014}.

We suggest that the reduction in deep focus is due to AI tools working for seconds, minutes, or hours to complete tasks given by the engineer. This results in a major task switch by the engineer and occasionally results in idle time. Both task switching and taking breaks interrupts focus resulting in cognitive overhead effort. So not only do engineers lose the reward of intense, deep focus, but they also are required to switch tasks more often.

\subsubsection{I have control over the code design and architecture; I lose track of what a piece of code is supposed to do.} \label{sec:control-over-design} This section is similar to results discussed in Section~\ref{sec:results-trust} which talks about engineers putting more and more trust into code written by AI. As engineers' work becomes more and more abstracted from design and even architectural details of code, their awareness of and control over design decisions drops. This is inevitable, but what we don't know is what it means for an engineer's affect; whether this results in a reduction of a sense of ownership and the associated positive consequences~\cite{pierce2003state} is unknown.

\subsection{Activities with no detectable difference}
In this section, we discuss some of the activities that didn't yield statistically detectable differences between AI and no-AI programming.

\subsubsection{I come up with new algorithms.} This result indicates that, while developers offload design decisions to AI (see Section~\ref{sec:control-over-design} and~\cite{zamfirescu2025beyond}), it appears that engineers may remain closely engaged in the process of writing algorithms, or at least they don't feel as strongly that AI is doing it for them. What, exactly, is different about designing software and designing algorithms, and whether developers are retaining the task of algorithm design because they enjoy or because AI isn't as good as algorithm development is unclear.

\subsubsection{I find and fix bugs in code.} This was another surprise. While developers use the debugger less when using AI (see Section~\ref{sec:less-debugger}), they may still find and fix bugs about the same.

\subsubsection{I repeat the same type of programming task many times.} A theme in software development is avoiding doing the same thing twice, yet even with the automation possibilities of GenAI tools, engineers still report performing repetitive tasks. We suggest that this is because, similar to writing software without the use of AI, incentives to write reusable code fail to be competitive against time pressures to deliver software, and even when using AI, writing reusable code takes effort. In fact, the problem may be exacerbated by the ease with which engineers can have AI write fresh, new code.

\subsubsection{I create unit tests manually to verify correctness.} For as much as they use AI to automate mundane tasks, software engineers appear unwilling to give up control of unit tests. This may indicate a continued distrust of code generated by AI, but it is somewhat surprising considering that enjoyment of the routine task of writing unit tests is mixed~\cite{daka2014unit}. Indeed, Sergeyuk et al.~\cite{sergeyuk2025using} found that engineers did desire to offload test writing to AI, with $61\%$ of respondents stating that they used AI to generate tests. However, only $32\%$ stated that they used AI to ``[find] untested areas of code.'' This is consistent with our results and highlights that ``people are less willing to use AI in stages requiring precision''~\cite{sergeyuk2025using}.

\subsubsection{I rush to complete programming work due to deadlines.} Apparently engineers still work against deadlines, even when working with AI assistance.

\subsection{RQ2: \rqtwo{}}
We refer to the association between programming experience and the magnitude of perceived differences as the \emph{experience gradient}. Every single activity and experience but one (``I receive feedback from instructors/TAs/supervisors regarding code'') that can be statistically shown to differ when using AI compared to not using AI has a difference that is even stronger among engineers with five or more years of experience. This could be due to multiple causes: one is that engineers experienced in writing software without AI assistance experience a more jarring change when transitioning to AI. Another reason could be that younger engineers' experience with AI-assisted development could be more gradual and subtle, graying the line between developing with and without AI, whereas more experienced developers may have made a more sudden and deliberate transition.

A third reason that experienced engineers' perceived difference in activities when and when not using AI is stronger than less experienced developers could be that experienced engineers may have biases and expectations that are influencing their responses. This last reason may be supported by our result that the nine survey respondents that had never used AI had a stronger difference in AI/no-AI activity and experience than those that had used AI (see Section~\ref{sec:nonusers}).

\section{Threats to Validity}
\label{sec:threats}

\textbf{Internal and construct validity.} All measures are self-reports, not observed behavior. AI and No-AI ratings were given side by side (AI first), and No-AI ratings are partly retrospective, so contrast or nostalgia effects are possible despite the instruction that the ratings are independent. The items are single, unvalidated statements introduced as ``how much each task is correlated to AI and No-AI based programming,'' so a rating may reflect frequency, applicability, or preference, and ``programming with AI'' was not defined (e.g., completion, chat, or agentic tools). Experience is confounded with role and cohort (8 of 9 students have $<$5 years; all researchers and educators $\geq$5 years), so RQ2 shows an association, not an effect of experience; the engineers-only analysis (Section~\ref{sec:robustness}) partly addresses role.

\textbf{External validity.} Our convenience sample came from one U.S.\ university's mailing list and several online developer communities, whose members are likely more engaged with, and more opinionated about, AI tools than the average programmer; we did not record how respondents found the survey, so recruitment channel may be confounded with role and experience. The sample is small ($n = 48$; 14 vs.\ 34 for experience groups), and the excluded non-users differ markedly (Section~\ref{sec:nonusers}), so the findings do not describe non-adopters, other tools, or other contexts.

\textbf{Statistical and data validity.} Item-level tests have low power, so non-significant items are not evidence of no change, and the $\pm 0.5$ equivalence margin is a judgment call. RQ2 and its five-year cutoff were specified post hoc, so RQ2 is exploratory. Because responses were anonymous, role, experience, and duplicate submissions could not be verified.

\section{Conclusions}
This paper is a contribution to the effort to define what the ``next generation'' software engineer looks like. Traditional software development relies on searching documentation, using a debugger, code reviews, and designing code, all of which we have found to be reduced in importance when using AI. Understanding that these may not be common activities moving forward will be critical in our preparation of future professionals as well as in continued professional development of existing software engineers. The change in experiential aspects of programming could also have a significant impact on the software engineering career. We found that when programming with AI, engineers tend to talk with peers less often, enter a deep focus state less often, and feel less in control of the code. As a result, the sense of community, absorption into the coding experience, and sense of ownership may be impacted, changing the engineering experience in fundamental ways. Together, these are the \emph{different craft} results: larger perceived differences in the programming activities and experiences surrounding code production, alongside relatively small mean differences in core code-production activities.

We need to be thinking about more than just the changing engineering activities---the entire affective experience of writing software is shifting. This could impact engineers' sense of belonging and identity as well as enjoyment and may attract a different type of person to the software engineering community. To support the new wave of software engineers we will need updated recruitment strategies, training curricula, reward systems, and company cultures. While there isn't yet any serious reason to believe that software development will move away from being a technical activity (indeed, we found that developers still engineer unit tests and algorithms when using AI), it is becoming clear that engineers are becoming more managers of agents and are putting some measure of trust in the agents' work.

A subtle but particularly important finding is the experience gradient: differences in perceived activity and experience between programming with and without AI are larger for more seasoned software engineers. Whether these perspectives are based solely on their experiences or are influenced by preconceptions about AI is less important than the fact that the engineers feel their careers shifting beneath them. Certainly many software developers will embrace the change, but others, such as those who depend on the rush of being in flow, who enjoy learning new programming languages, and those who value the social interactions of discussing code with other humans may become disenchanted with software development. Whether this alienation is reconcilable or not is an open question that will depend on further characterization of what, exactly, engineers enjoy and whether engineering with AI has a place for new forms of the most rewarding activities and experiences.

\section{Acknowledgments}
\textbf{Use of generative AI.} Claude (Anthropic) AI was used in preparing
this paper. It was used to draft and revise text in the Introduction, Related Work, and Methods sections. It was also used to check claims against cited papers and prepare the BibTeX file. The authors designed the study, collected and interpreted the data, and reviewed, edited, and verified all AI-assisted text, code, and references. The authors take full responsibility for the content of this paper.
\section{Data Availability}
Anonymized survey data, analysis notebooks, the survey instrument, and more detailed item-to-literature mapping are available at \url{https://osf.io/gkw4v/overview?view_only=a313ac8db7254464a666be4813d1effd}. We will make these materials public upon acceptance.
\bibliographystyle{ACM-Reference-Format}
\bibliography{actual_paper}

\end{document}